\documentclass[a4paper]{article}

\usepackage{INTERSPEECH2021}
\usepackage{hyperref}
\usepackage{pgfplots}
\usepackage{pgfplots}
\usepackage{amsmath,subcaption}
\DeclareUnicodeCharacter{2212}{-} 
\usepgfplotslibrary{groupplots,dateplot}
\usetikzlibrary{patterns,shapes.arrows}
\pgfplotsset{compat=newest}
\usepackage{hyperref}
\usepackage{pdfpages}
\usepackage[textsize=tiny]{todonotes}
\usepackage{enumitem,kantlipsum}
\title{pMCT: Patched Multi-Condition Training for Robust Speech Recognition}
\name{Pablo Peso Parada, Agnieszka Dobrowolska, Karthikeyan Saravanan, Mete Ozay}
\address{
  Samsung Research, UK}
\email{\{p.parada, aga.dobrowol, k1.saravanan, m.ozay\}@samsung.com}

\newcommand{\probnoise}{p_{n}} % define here to easily change
\newcommand{\probreverb}{p_{r}} % define here to easily change

\newlength\figH
\newlength\figW
\pgfplotsset{/pgfplots/group/.cd,
    horizontal sep= 1.5cm,
    vertical sep= 1.5cm
}
\pgfplotsset{every tick label/.append style={font=\scriptsize}}

\begin{document}

\maketitle
% 

% ====================================================================

\begin{abstract}
We propose a novel Patched Multi-Condition Training (pMCT) method for robust Automatic Speech Recognition (ASR). pMCT employs Multi-condition Audio Modification and Patching  (MAMP) via mixing {\it patches} of the same utterance  extracted from clean and distorted speech. Training using patch-modified signals improves robustness of models in noisy reverberant scenarios. Our proposed pMCT is evaluated on the LibriSpeech dataset showing improvement over using vanilla Multi-Condition Training (MCT). For analyses on robust ASR, we employed pMCT  on the VOiCES dataset which is a noisy reverberant dataset created using utterances from LibriSpeech. In the analyses,  pMCT achieves 23.1\% relative WER reduction compared to the MCT.%\todo[inline]{Needs mentioning manifold}

\end{abstract}
\noindent\textbf{Index Terms}: robust speech recognition, data augmentation.

\section{Introduction\label{sec:intro}}

% Introduction
With the rising integration of voice search and interaction systems in products such as mobile devices and home appliances, there is an increasing need for robust Automatic Speech Recognition (ASR) systems. One of the challenges of training ASR models arises from the mismatch between training and test conditions; models trained on curated close-talk training datasets do not generalize well to real-world conditions with additive noise, channel distortion and reverberation  \cite{DBLP:journals/interspeech/abs-2106-03419}.

 Considering the number and variety of acoustic conditions observed in real-world scenarios, acquisition of sufficiently diverse training data is not trivial.  Hence, data augmentation (DA) is commonly utilized to diversify and enlarge the training corpus, acting as a regularizer to prevent overfitting and thus increasing overall model robustness. A number of approaches based on signal processing have been proposed. Time-domain augmentations include modifying the sampling rate by a small factor, such as $\pm 10\%$, which changes pitch and duration of the audio \cite{DBLP:conf/interspeech/KoPPK15}.
%
%
% Vocal Track Length Perturbation
Some approaches focus on augmenting speaker characteristics, e.g., Vocal Tract Length Perturbation (VLTP) \cite{jaitly2013vocal,kim2019improved}.
% modifies the Vocal Track Length -- a speech characteristic which varies from person to person.
%
% Speed perturbation
Speed perturbation \cite{DBLP:conf/interspeech/KoPPK15}, a widely used augmentation, can be seen as emulating tempo perturbation and VTLP. 
%
%
% Volume perturbation 
As the variance in audio volume can be low in curated datasets, volume perturbation \cite{DBLP:conf/interspeech/PeddintiPK15} has been used to randomly scale the training data. % emulating mean shifts in the MFCC domain

 SpecAugment \cite{DBLP:conf/interspeech/ParkCZCZCL19} acts by randomly masking chunks of time (time masking) or frequency channels (frequency masking) on spectrograms, and applying a time warping transformation. Recently, a number of variants of SpecAugment have been proposed.
SpliceOut \cite{DBLP:journals/corr/abs-2110-00046} removes selected time blocks entirely from audio, and concatenates  the remaining parts together.
SpecAugment++ \cite{DBLP:journals/corr/abs-2103-16858} applies frequency and time masking not only to the input spectrogram but also to latent spaces  of neural networks (NNs). In addition, SpecAugment++ proposes filling the masked out areas with the time frames and frequency channels of other samples within the mini-batch. Similarly, SpecSwap \cite{DBLP:conf/interspeech/SongWH0M20} swaps time blocks and frequency blocks. However, the interchanged sections originate from the same audio. Some works \cite{DBLP:journals/spl/CarrBBTZ21} propose learning to reorder the shuffled spectrogram patches, utilizing a fully-differentiable permutation loss.
% 
%
% Frequency domain
Systematic omission of frequency channels of the input spectrogram has been studied in \cite{DBLP:journals/prl/KovacsTCG17,DBLP:conf/specom/TothKC18}.

MixSpeech \cite{ DBLP:conf/icassp/MengX00Q021}, an adaptation of Mixup \cite{DBLP:conf/iclr/ZhangCDL18} for ASR, %first reference is to mixup, second to mixspeech
and Between-Class (BC) learning \cite{DBLP:conf/iclr/TokozumeUH18} generate new training samples by interpolating multiple audio samples. A generalization of BC, SpeechMix \cite{speechMix}, applies interpolations on latent spaces. %While earlier-mentioned approaches apply label-preserving transformations, MixSpeech, BC and SpeechMix all alter the labels, which has been shown to result in training instabilities in some cases. 

Another family of DA approaches targets improving performance in noisy and reverberant scenarios.
%
%
%\textbf{Additive Noise:} 
A common approach for diversifying the training corpus is to add noise from another source to the original waveform \cite{DBLP:journals/corr/abs-1301-3605,DBLP:journals/corr/HannunCCCDEPSSCN14}, sampling the SNR levels from a random distribution. ImportantAug \cite{trinh2021} learns an importance map for the audio and only adds noise to unimportant regions.  Generative Adversarial Networks are used to match the noisy test conditions in \cite{DBLP:conf/icassp/SriramJGS18,DBLP:journals/corr/abs-2106-03419}. Multi-Condition Training (MCT)~\cite{kim2017,ko2017}, also referred to as Multistyle TRaining (MTR) \cite{park2020}, combines speech distortion with Room Impulse Response (RIR) and additive noise to create robust ASR models~\cite{kim2017,kim2021}.

Our contributions can be summarized as follows:
%\begin{itemize}[leftmargin=*]

\noindent
	$\bullet$ We propose a simple yet effective MCT method called Patched Multi-Condition Training (pMCT) to improve ASR accuracy, especially for noisy reverberant conditions. pMCT employs a novel method, called Multi-condition Audio Modification and Patching (MAMP), which mixes same utterance patches extracted from clean and distorted speech (Fig.~\ref{fig:pMCT}).

	\noindent
	$\bullet$ Our MAMP can be combined with additional data augmentation methods such as SpecAugment \cite{DBLP:conf/interspeech/ParkCZCZCL19} to improve accuracy of pMCT. We found that employing pMCT with SpecAugment further improves WER on both clean (up to 0.8\% relative WER reduction) and noisy (up to 22.9\% relative WER reduction) datasets.
%\end{itemize}

The paper is organised as follows: Section~\ref{sec:method} presents the method proposed in this work. The evaluation setup and the results obtained are detailed in  Section \ref{sec:eval} and Section \ref{sec:results}. Finally, the conclusions are drawn in Section \ref{sec:conclusions}.

% ====================================================================
\vspace{-0.2cm}
\section{Patched Multi-Condition Training for Robust Speech Recognition\label{sec:method}}

In this section, we describe our proposed patched multi-condition training pipeline. We employ the MAMP method in pMCT to improve accuracy of ASR models in real-world ASR use-cases, e.g., in far field set-ups where reverberation and noise significantly distort the speech signal. In the next subsection, we introduce the theoretical motivation of the proposed method. 
\vspace{-0.2cm}

\subsection{Data Augmentation with Audio Patches}

ASR algorithms  rely on prior knowledge via inductive biases e.g., by using deep neural networks (DNNs) when the  distribution
$\mathcal{P}$ of speech signals $x(n)$ is unknown. 
%Similarly, data augmentation (DA) strategies exploit known invariants of ASR models, e.g. being invariant to phone-preserving transformations \cite{barreda2021}.
Similarly, some DA strategies exploit prior knowledge about what is in support $\sigma(\mathcal{P})$ of $\mathcal{P}$ such as SpecAugment. However, particularly in real-world applications such as ASR in noisy and reverberant conditions, the prior information about noise and reverberation may not be available in the $\sigma(\mathcal{P})$. Therefore, the MCT approach aims to exploit prior knowledge about what is not in the $\sigma(\mathcal{P})$ while training DNNs. 

This prior knowledge is often available for audio data. Specifically, we assume that (1) there exists  an alternative
distribution $\mathcal{Q}$ with a different support $\sigma(\mathcal{Q})$; and (2) we have access to a procedure to
efficiently sample from $\mathcal{Q}$. We emphasize that $\mathcal{Q}$ is not required to be explicitly defined (e.g., through an explicit density). This alternative distribution  $\mathcal{Q}$ may be implicitly defined by a dataset collected in these scenarios, or by a procedure that transforms samples from $\mathcal{P}$ into ones from $\mathcal{Q}$ by suitably altering their structure, such as by audio modification of $x(n)$ to obtain a distorted speech signal $y(n)$.

According to the manifold hypothesis \cite{Testing}, utterances  lie on a $d$ dimensional manifold $\mathcal{M}$ of the $D\gg d $ dimensional ambient (signal) space $\mathcal{N}$ containing both clean ($x(n)$) and distorted ($y(n)$) signals. More precisely, since manifolds can be approximated by locally Euclidean patches in small neighborhoods,  phonemes and utterances reside on audio patches \cite{vaz}. Thereby, support $\sigma({\mathcal{P}})$ resides on $\mathcal{M}$.  Manifold learning methods \cite{vaz,svd} assume that speech does not exist in the remaining $D-d$ dimensions. Then, they aim to project noisy data onto noiseless subspaces, such as to the manifolds formed by the eigenvectors corresponding to the $d$ largest singular values of noisy data \cite{svd}.  

We do not perform  computationally complex projections from data space or the support $\sigma(\mathcal{Q})$ on $\mathcal{M}$ for noise removal. Instead, we consider that highly entangled noise and different parts of speech can be untangled across utterance manifolds at different layers of DNNs to achieve invariant ASR \cite{unt}. In addition, DA is informative if the support of distribution of data generated by DA on $\mathcal{M}$ is close  to $\sigma(\mathcal{Q})$ \cite{sinha2021negative}. These noisy samples will provide information on the relationship between $\sigma(\mathcal{P})$ and $\sigma(\mathcal{Q})$, which we will show is helpful in ASR. This suggests that many DA strategies, such as sampling from a noise or reverberant set, exist. 

We propose employing the inductive bias by training DNNs with DA on audio patches approximating utterance manifolds to improve robustness of ASR models against acoustic distortions. Untangled representations learned at different layers of DNNs were explored using attention maps \cite{YangLL20,shim2022understanding}. To elucidate how different representations are projected on feature spaces at different layers of DNNs, we analyze their eigenspectrum in Section~\ref{sec:self-attention}.

%In our ASR models, the speech signals are processed using NNs that are good at processing local features but not necessarily global features. Therefore, we consider DA examples that can preserve local features (``informative'') and break global features, so that it forces the NNs to learn global features.

\begin{figure}[t]
	\centering
	\input{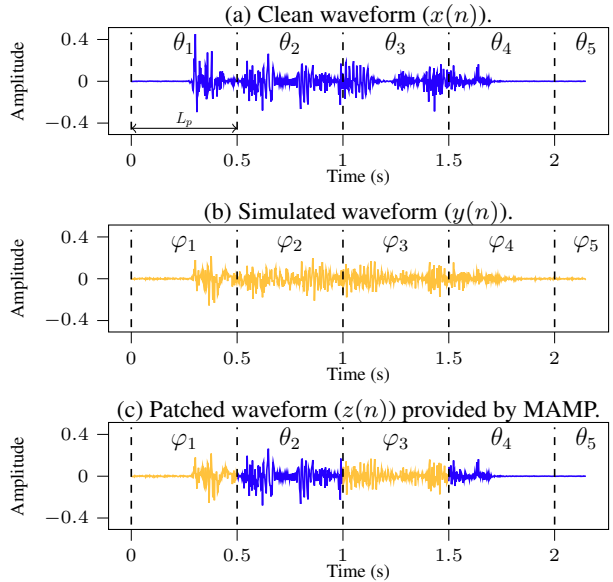}
	\vspace{-15pt}
	\caption{(a) Waveform of a  clean speech signal $x(n)$ of length $L$ split into  patches $\theta_p$ with length $L_p$ for $p=1,2,\ldots,5$. (b) Noisy reverberant waveform $y(n)$, simulated using \eqref{eqn:y}, split into patches $\{\varphi_p\}_{p=1}^5$. (c) Augmented signal $z(n)$ comprises segments from $x(n)$ and $y(n)$. Vertical dashed lines indicate the size of each patch, $L_p=500~ms$.}
	\vspace{-0.5cm}
	\label{fig:pMCT}
\end{figure}

\subsection{Multi-condition Audio Modification and Patching (MAMP)}

Patches are extracted from audio signals and mixed as follows:
\begin{enumerate}[leftmargin=*]
	\item \textbf{Modification step:} A clean speech signal $x(n)$ (Fig.\ref{fig:pMCT}(a)) is modified to obtain $y(n)$ (Fig.\ref{fig:pMCT}(b)) following the data augmentation employed in MCT. As described in \cite{kim2017,ko2017,park2020,kim2021}, MCT performs data augmentation by convolving a speech signal $x(n)$ with a Room Impulse Response (RIR) $h(n)$  to introduce reverberation followed by adding a noise source $\nu(n)$ to generate the simulated far-field speech $y(n)$
\vspace{-0.2cm}
\begin{equation}
	y(n) = \sum_{m=0}^{M-1} h(m)x(n-m) + \nu(n)
	%y(n) =   h(n) \star x(n) + \nu(n)
	\label{eqn:y}
\vspace{-0.15cm}
\end{equation}

where $M$ is the effective length of $h(m)$ and $n$ is the time index. Applying this transformation {\it on-the-fly} generates highly diverse training scenarios without increasing storage.
	
	\item \textbf{Patch extraction step:} The signals $x(n)$ and $y(n)$ are split into patches $\theta_p$ and $\varphi_p$, respectively of size $L_p$, as depicted by the vertical dashed lines in Fig.\ref{fig:pMCT}, where ${\sum_{p=1}^P L_p = L}$ and $L$ is the length of $x(n)$. Note that $L_P \leq L_1$, since $L$ may not be multiple of the patch length. Thereby, we obtain clean patches $\lbrace\theta_p\rbrace_{p=1}^P$  and distorted patches $\lbrace\varphi_p\rbrace_{p=1}^P$ where $P$ is the total number of patches.
	
	\item \textbf{Patch mixing step:} The MAMP signal $z(n)$ is obtained by determining a patch $\phi_p$ as a clean patch $\phi_p:= \theta_p$ with a probability  $\pi$, or as a distorted patch $\phi_p:=\varphi_p$ with the probability $1-\pi$, where $:=$ denotes the assignment operator. %Then, the selected patch is appended to the set  $\mathcal{F}=\{{\phi_p\}}_{p=1}^P$ where 
	Each element $\phi_p$ is the $p^{th}$ patch of the signal $z(n)$ as shown in Fig.\ref{fig:pMCT}(c). The delay of the RIR's direct path is removed in order to align the $x(n)$ and $y(n)$ so that they can be mixed. 
	
\end{enumerate}

\subsection{Patched Multi-Condition Training (pMCT)}
Finally,  pMCT is performed following the standard MCT training recipe \cite{ko2017}, but employing the MAMP to augment the data instead of only utilizing Eq. \eqref{eqn:y}.

\section{Evaluation of pMCT in ASR Tasks \label{sec:eval}}

\subsection{Experimental setup}
In the experiments, we implemented ASR models using the transformer recipe in SpeechBrain \cite{speechbrain} for LibriSpeech\footnote{\scriptsize{\href{https://github.com/speechbrain/speechbrain/blob/develop/recipes/LibriSpeech/ASR/transformer}{github.com/speechbrain/speechbrain/blob/develop/recipes/LibriSpeech/ ASR/transformer}}}. 
The recipe implements an end-to-end transformer ASR architecture with a Conformer encoder \cite{conformer}. The Conformer configuration follows the Conformer (S) described in \cite{conformer}, and we used a transformer language model. The loss is computed using a weighted average of the CTC loss and KL-divergence with label smoothing set to 0.15. The models are trained for 200 epochs and the best model is selected based on the Word Error Rate (WER) achieved in validation set. All the experiments implemented using the 100h dataset are reported using a beam search of size 1 to speed up the analyses, while the final results on the 960h dataset are obtained  using a beam search of size 10.

\subsection{Datasets}
 
\textbf{LibriSpeech Corpus:} LibriSpeech \cite{librispeech} comprises a 960h dataset of English utterances, sampled at 16kHz. In this work, we use either {\it train-clean-100} -- a 100h subset, or the entire 960h dataset for training. We evaluate models using the validation set {\it dev-clean} as well as the test sets {\it test-clean} and {\it test-other}. Model accuracy is reported in terms of WER. 

\noindent
\textbf{VOiCES:} VOiCES \cite{voices} is a dataset created by recording LibriSpeech utterances played in acoustically-challenging conditions. Using this dataset for evaluation  allows assessing the  accuracy of the proposed methods in realistic and noisy reverberant environments. The evaluation set for ASR comprises of 20h of data originating from 98 different speakers.

\noindent
\textbf{Datasets used for DA:} Two main sets are used for DA: (1) comprising RIRs, and (2) with noise recordings.
We use a set of RIRs gathered from 3 datasets \cite{ko2017}, giving rise to 325 different RIRs. The level of reverberation ranges from -2~dB to 40~dB in terms of C50 \cite{C50}.
The noise dataset uses a set of 843 noise samples obtained from the Freesound portion of the MUSAN corpus \cite{musan}. The Signal-to-Noise Ratio (SNR) is randomly sampled from a uniform distribution between 0-30~dB.

\section{Experimental Analyses \label{sec:results}}
 
\subsection{Ablation Studies}
\label{ablation}
In this section, we provide ablation studies on \emph{dev-clean} set to analyze the effect of using different patch selection probabilities, $\pi$, patch lengths $L$, and SpecAugment policies on the accuracy of models. Experiments to find the best hyperparameters were performed on the 100h subset of LibriSpeech.

\subsubsection{Analyses of the Effect of Patch Selection Probability $\pi$}

Firstly, we investigate how model accuracy varies as a function of the probability of a clean patch $\pi$ for pMCT in Table \ref{tab:pMCT:p_clean}.  In the table, {\it rand} indicates that the utterance is selected with a different probability value drawn from a uniform distribution. We find that best WER is achieved using $\pi=0.5$.

\vspace{-0.15cm}
\begin{table}[ht!]
  \caption{Results on the dev-clean set for different values of $\pi$.}\vspace*{-10pt}
  \label{tab:pMCT:p_clean}
  \centering
  \begin{tabular}{r|cccccc}
    \toprule
%     & \multicolumn{6}{c}{$\boldsymbol{\pi}$} \\
	 $\boldsymbol{\pi}$ & 0 & 0.25 & 0.5 & 0.75 & 1 & rand \\
%    \midrule
    \textbf{WER (\%)} & 10.53 & 10.04 & {\bf 9.76} & 9.84 & 10.91 & 9.98\\
    \bottomrule
  \end{tabular}
\end{table}
\vspace*{-0.5cm}

\subsubsection{Analyses of the Effect of Patch Size $L_p$}

Next, we investigate how varying patch size, $L_p$, affects training. Different sizes are tested as indicated in Table \ref{tab:pMCT:patch_size}, suggesting that the optimal patch size is between 1 and 1.5 seconds.
\vspace{-0.15cm}
\begin{table}[ht!]
  \caption{Results on the dev-clean set for different patch size  $L_p$.}\vspace*{-10pt}
  \label{tab:pMCT:patch_size}
  \centering
  \begin{tabular}{r|ccccc}
    \toprule
    \textbf{$\boldsymbol{L_p}$ (s)} & 0.5 & 0.75 & 1 & 1.5 & 2  \\
%    \midrule
    \textbf{WER (\%)} & 9.79 & 9.78 & {\bf 9.76 } & {\bf 9.76 } & 9.98 \\
    \bottomrule
  \end{tabular}
\end{table}
\vspace{-0.5cm}

\subsubsection{Analyses of the Effect of SpecAugment Policies}

%Since both, pMCT and SpecAugment, have been shown to improve model performance, it is important to explore the effect of combining these methods.
% Different experiments are carried out to analyse the interaction of pMCT with SpecAugment. 
SpecAugment is widely used to train ASR models due to its efficiency \cite{DBLP:conf/interspeech/ParkCZCZCL19,park2020}. In this section, we investigate whether the strength of SpecAugment should be modified when used in conjunction with pMCT.
To check this, we experimented with modifying  masking strength of Adaptive SpecAugment \cite{park2020}, as shown in Table \ref{tab:pMCT:SpecAugment}. {\it High} policy corresponds to using the LibriFullAdapt policy \cite{park2020}, whereas the \textit{Mid} and \textit{Low} policies are obtained by dividing the masking parameters, such as the number of masks and mask length, by 2 and 4, respectively. We find that when using SpecAugment together with pMCT, aggressiveness of SpecAugment should be lowered to achieve best accuracy. Hence, throughout this work, we use the {\it Mid} policy when using SpecAugment alongside pMCT, unless stated otherwise.

\vspace{-0.2cm}
\begin{table}[ht!]
  \caption{Comparison of WER on the dev-clean set for different SpecAugment settings applied in conjunction with pMCT.}\vspace*{-10pt}
  \label{tab:pMCT:SpecAugment}
  \centering
  \begin{tabular}{r|ccc}
    \toprule
    \textbf{Setting}& High & Mid & Low\\
    \textbf{WER (\%)} & 11.79 & {\bf 9.3} & 9.96\\
    \bottomrule
  \end{tabular}
\end{table}
\vspace{-0.5cm}

\subsubsection{Ablation for the MCT}
As a baseline for pMCT, we use MCT. Three settings are explored for the MCT method~\cite{ko2017}: probability $\probreverb$ of adding reverberation to the input signal, probability $\probnoise$ of adding noise to the signal  and the SpecAugment setting. For simplicity, we set $\probnoise = \probreverb$. Table \ref{tab:MCT:patch_size} shows that the lowest WER is achieved with $\probnoise=0.5$ and $\probreverb=0.5$, and this is the setting used for MCT for the remainder of the paper. Results obtained by employing SpecAugment on MCT are shown in Table \ref{tab:MCT:SpecAugment}, indicating that the best results are obtained for using the \textit{Mid} setting.

\vspace{-0.25cm}
\begin{table}[ht!]
  \caption{WER of models on the dev-clean of MCT with different probabilities applied to noise $\probnoise$ and reverberation $\probreverb$.}\vspace*{-10pt}
  \label{tab:MCT:patch_size}
  \centering
  \begin{tabular}{r|ccccc}
    \toprule
    $\boldsymbol{\probnoise} = \boldsymbol{\probreverb}$ & 0 & 0.25 & 0.5 & 0.75 & 1 \\ 
%    $\boldsymbol{\probreverb}$ & 0 & 0.25 & 0.5 & 0.75 & 1 \\
    %\midrule
       % \cline{2-6}
    \textbf{WER (\%)} & 10.91& 10.05& {\bf 10.00}& 10.01 & 10.56 \\
    \bottomrule
  \end{tabular}
\end{table}
\vspace{-0.35cm}

\vspace{-0.25cm}
\begin{table}[ht!]
  \caption{Comparison of WER of models on the dev-clean set for different SpecAugment settings applied on top of MCT.}\vspace*{-10pt}
  \label{tab:MCT:SpecAugment}
  \centering
  \begin{tabular}{r|ccc}
    \toprule
    \textbf{Policy} & High & Mid & Low \\
    \textbf{WER (\%)} & 10.16 & {\bf 9.25} & 9.59 \\
    \bottomrule
  \end{tabular}
\end{table}
\vspace{-0.5cm}

\subsubsection{Comparative Evaluation on the Test Datasets}

Table \ref{tab:100h} summarizes all the results and extends the evaluation of the methods to three other datasets: test-clean, test-other and VOiCES. The baseline in this case corresponds to training models with LibriFullAdapt policy for SpecAugment, but without using MCT or pMCT. The results show that pMCT with SpecAugment achieves best results on the majority of the datasets, with largest gains observed on VOiCES.

% TODO: Clear at this point which are the settings used for each of the methods?
\vspace{-0.25cm}
\begin{table}[ht!]
  \caption{Results for models trained on the 100h dataset.}\vspace*{-10pt}
  \label{tab:100h}
  \centering
  \begin{tabular}{ccccc}
    \toprule
    \textbf{Method} & \textbf{\begin{tabular}{@{}c@{}}dev-\\clean\end{tabular}} & \textbf{\begin{tabular}{@{}c@{}}test-\\clean\end{tabular}} & \textbf{\begin{tabular}{@{}c@{}}test-\\other\end{tabular}} & \textbf{VOiCES}\\
    \midrule
    Baseline & 9.42 & 9.56 & 24.51 & 75.07 \\
    MCT w/o specAug & 10.00 & 9.23 & 24.74 & 36.79 \\
    pMCT w/o specAug & 9.76 & 9.39 & 25.09 & 31.58 \\
    MCT w/ specAug & {\bf 9.25} & 9.50 & 24.29 & 36.19 \\
    pMCT w/ specAug & 9.30 & {\bf 9.15} & {\bf 23.21} & {\bf 27.97} \\
    \bottomrule
  \end{tabular}
\end{table}
\vspace{-0.5cm}

\subsection{Analyses for Training Models on Large  Datasets}
In this section, we repeat the experiments using the full 960h from LibriSpeech to train the models (Table \ref{tab:960h}). As in the case for Table \ref{tab:100h}, the baseline in Table \ref{tab:960h} refers to training  models with LibriFullAdapt policy for SpecAugment but without using MCT or pMCT.

 As shown in Table \ref{tab:960h}, pMCT achieves the lowest WER on most of the datasets, with a 23.1\% relative improvement on VOiCES in comparison with MCT when SpecAugment is used for both.
 The results also highlight the importance of multi-conditional training for robust ASR -- even in the worst case (MCT w/o SpecAugment) more than a 50\% relative WER reduction is achieved on VOiCES.  SpecAugment also provides additional improvement when applied after MCT or pMCT as shown in Table \ref{tab:960h}. 

% TODO: Why p2dMCT doesn't work that well????
\vspace{-0.25cm}
\begin{table}[ht!]
  \caption{Results for models trained on the 960h dataset. }\vspace*{-10pt}
  \label{tab:960h}
  \centering
  \begin{tabular}{rcccc}
    \toprule
    \textbf{Method} & \textbf{\begin{tabular}{@{}c@{}}dev-\\clean\end{tabular}} & \textbf{\begin{tabular}{@{}c@{}}test-\\clean\end{tabular}} & \textbf{\begin{tabular}{@{}c@{}}test-\\other\end{tabular}} & \textbf{VOiCES}\\
    \midrule
    Baseline & 2.59 & 2.74 & 6.51 & 33.30 \\
    MCT w/o specAug & 2.43 & 2.72 & 7.07 & 14.85 \\
    pMCT w/o specAug & 2.44 & 2.67 & 6.90 & 12.83  \\
    MCT w/ specAug & {\bf 2.42} & 2.65 & 6.59 & 12.86\\
    pMCT w/ specAug & {\bf 2.42} & {\bf 2.41} & {\bf 6.36} & {\bf 9.89} \\
    \bottomrule
  \end{tabular}
\end{table}
\vspace{-0.15cm}

The audio samples of the VOiCES dataset were recorded in four rooms, $\{R_i\}_{i=1}^4$. We examined the WER on each of the rooms for the models with the best WER on the \emph{dev-clean} dataset. The results given in Table  \ref{tab:VOiCES_breakdown_room} show that pMCT outperforms MCT for all rooms. It is also clear that the models perform worse for larger rooms ($R_3$ and $R_4$) due to increased reverberation in these setups.

\vspace{-0.25cm}
\begin{table}[ht!]
  \caption{Analyses of the models for the four different rooms included in evaluation dataset of VOiCES.}\vspace*{-10pt}
  \label{tab:VOiCES_breakdown_room}
  \centering
  \begin{tabular}{rcccc}
    \toprule
    \textbf{Method} & \textbf{$R_1$} & \textbf{$R_2$} &\textbf{$R_3$} &\textbf{$R_4$} \\
    \midrule
    MCT w/ specAug  & 5.70 & 5.68 & 24.13 & 15.19 \\
    pMCT w/ specAug & {\bf 4.74} & {\bf 5.18} & {\bf 15.34} & {\bf 13.78} \\
    \bottomrule
  \end{tabular}
\end{table}
\vspace{-0.05cm}

Table \ref{tab:VOiCES_breakdown_noise} shows the WER provided by these best MCT and pMCT models for different noise types in the evaluation dataset of the VOiCES. In general, pMCT outperforms MCT for all types of noises. Moreover, babble noise tends to be the most challenging noise due to the similarity with the target speech.
\vspace{-0.15cm}
\begin{table}[ht!]
  \caption{Analyses of the models for 4 noise types; no noise (\textbf{none}),  music (\textbf{music}), television  (\textbf{tele}) and babble (\textbf{babb}).}\vspace*{-10pt}
  \label{tab:VOiCES_breakdown_noise}
  \centering
  \begin{tabular}{rcccc}
    \toprule
    \textbf{Method} & \textbf{none} & \textbf{music} &\textbf{tele} &\textbf{babb} \\
    \midrule
    MCT w/ specAug  & 15.33 & 7.94 & 7.93 & 15.84 \\
    pMCT w/ specAug & {\bf 10.02} & {\bf 6.67} & {\bf 7.17} & {\bf 14.81} \\
    \bottomrule
  \end{tabular}
\end{table}
\vspace{-0.5cm}

 \subsection{Analyses of Self-attention in MCT and pMCT} \label{sec:self-attention}
The self-attention mechanism in the transformer ASR architectures is used to relate different audio frames to extract a representation of the input sequence \cite{NIPS2017_3f5ee243}. In order to analyse the differences of the self-attention maps learned by models trained with MCT and pMCT, we use the covariance self-attention matrix \cite{bhojanapalli2021eigen} which captures interaction across different inputs, and compute the average skewness of their eigenvalues:
\vspace{-0.2cm}
\begin{equation}
\label{Eq:skew}
    \boldsymbol{S} = \mathbb{E}_{y\in\mathcal{D}}\left[ \frac{1}{L\cdot N}\sum_{l\in[L],h\in[H]} s\left(\mathbf{\lambda}^{l,h}_y \right)\right]
\end{equation}
where $y$ is the input utterance from the dataset $\mathcal{D}$, $s$ is the skewness function, $L$ is the total number of layers, $H$ is the total number of heads, $\mathbf{\lambda}^{l,h}_y$ is the vector comprising the absolute values of eigenvalues of $A_y^{l,h}$ ${A^{l,h}_y}^\top$ sorted decrementally, $A_y^{l,h}$ is the self-attention map of the $l^{th}$ layer and the  $h^{th}$ head. The  skewness for pMCT, $\boldsymbol{S}_p$, is 10\% to 11\% lower compared to the MCT, $\boldsymbol{S}_m$, in all datasets as shown in Table \ref{tab:skew}.
This indicates a reduction in the largest eigenvalues and also higher variations of the attention scores, and thus more connections between input sequence elements are achieved for pMCT using MAMP which helps to improve the ASR performance.

\begin{table}[t!]
  \caption{Summary of the skewness drop of pMCT models compared to MCT models measured as $(\boldsymbol{S}_m - \boldsymbol{S}_p)/\boldsymbol{S}_m$.}\vspace*{-10pt}
  \label{tab:skew}
  \centering
  \begin{tabular}{cccc}
    \toprule
    \textbf{dev-clean} & \textbf{test-clean} & \textbf{test-other} & \textbf{VOiCES}\\
    \midrule
     0.113 & 0.115 & 0.109 & 0.104 \\
    \bottomrule
  \end{tabular}
\vspace{-0.5cm}
\end{table}

\subsection{Comparison of MCT and pMCT employed on State-of-the-art Robust ASR Models}

To examine the benefits of using pMCT  in other methods proposed for robust ASR \cite{park2020,kim2021}, we employed pMCT to train ASR models and compared the results with the MCT counterpart. In \cite{park2020}, MCT is only applied to the 20\% of the data, whereas in \cite{kim2021}, MCT is applied in conjunction with VTLP.

Table \ref{tab:MTR960h} shows the WER achieved by using pMCT instead of using MCT in the method proposed in \cite{park2020}. The results indicate that using pMCT improves the WER over the MCT variant of the method. 
One of the reasons of observing the small difference between the WER of pMCT and MCT is that the multi-condition augmentation is only applied to 20\% of the training data leaving the remaining 80\% of the data clean. Thus, the difference in the final training data is reduced compared to applying multi-condition augmentation to the entire training set.

\vspace{-0.25cm}
\begin{table}[ht!]
  \caption{Comparison of WER (\%) of models trained on 960h using the method proposed in \cite{park2020} with and without pMCT.}\vspace*{-10pt}
  \label{tab:MTR960h}
  \centering
  \begin{tabular}{rcccc}
    \toprule
    \textbf{Method} & \textbf{\begin{tabular}{@{}c@{}}dev-\\clean\end{tabular}} & \textbf{\begin{tabular}{@{}c@{}}test-\\clean\end{tabular}} & \textbf{\begin{tabular}{@{}c@{}}test-\\other\end{tabular}} & \textbf{VOiCES}\\
    \midrule
    \cite{park2020} w/ MCT & 2.55 & 2.73 & 6.35 & 13.15 \\
    \cite{park2020} w/ pMCT & {\bf 2.32} & {\bf 2.5} & {\bf 6.3} & {\bf 13.09} \\
    \bottomrule
  \end{tabular}
\end{table}
\vspace{-0.25cm}

Table \ref{tab:VTLP960h} shows the WER achieved using the VTLP-based method proposed in \cite{kim2021}. The results indicate that the pMCT outperforms the MCT in terms of average WER.

% TODO: Make sure this approach VTLP + AS is explained in the introduction
\vspace{-0.25cm}
\begin{table}[ht!]
  \caption{Comparison of WER (\%) of models trained on 960h using the method proposed in \cite{kim2021} with and without pMCT.}\vspace*{-10pt}
  \label{tab:VTLP960h}
  \centering
  \begin{tabular}{rcccc}
    \toprule
    \textbf{Method} & \textbf{\begin{tabular}{@{}c@{}}dev-\\clean\end{tabular}} & \textbf{\begin{tabular}{@{}c@{}}test-\\clean\end{tabular}} & \textbf{\begin{tabular}{@{}c@{}}test-\\other\end{tabular}} & \textbf{VOiCES}\\
    \midrule
    \cite{kim2021} w/ MCT & {\bf 2.43}  & {\bf 2.54}  &  6.97 &  15.31 \\
    \cite{kim2021} w/ pMCT & {\bf 2.43}  & 2.64  & {\bf 6.86} & {\bf 11.15} \\
    \bottomrule
  \end{tabular}
\end{table}
\vspace{-0.5cm}

\section{Conclusions\label{sec:conclusions}}
We have presented a data augmentation method called pMCT to improve accuracy of ASR models,  which enhances MCT by employing patch modifications.
The proposed method is simple to implement, and introduces negligible computational overhead  without requiring additional data compared to MCT.
Experimental analyses on LibriSpeech and VOiCES show that pMCT outperforms MCT in both clean and noisy reveberant conditions.
The improvement is especially significant on the VOiCES dataset, achieving a relative WER reduction of 23.1\%.
An improvement of pMCT over MCT is also observed when incorporating pMCT into other training pipelines and robust ASR models \cite{park2020,kim2021} which traditionally employ MCT.
Lastly, we find that combining pMCT with other augmentation schemes, such as SpecAugment, achieves further improvements.  

\clearpage
\bibliographystyle{IEEEtran}

\bibliography{mybib.bib}

\end{document}